\documentclass{pasj00}

\begin{document}
\SetRunningHead{Matsumura et al.}{Interstellar Polarization and Temperature}

\title{Correlation between Interstellar Polarization and Dust Temperature: 
Alignment of Grains by Radiative Torques is Ubiquitous? }

%

%
\author{%
  Masafumi \textsc{Matsumura,}\altaffilmark{1}
  Youko   \textsc{Kameura,}\altaffilmark{1}
   \thanks{Present Address: Akita Junior High School, Magodai-machi 72, Kumamoto 861-5254 
           and Rikigo Junior High School, Sima-machi 5-8-1, Kumamoto 861-4133}
  Koji S. \textsc{Kawabata,}\altaffilmark{2}
  Hiroshi \textsc{Akitaya,}\altaffilmark{2}
  Mizuki  \textsc{Isogai,}\altaffilmark{3}
  \and 
  Munezo  \textsc{Seki}\altaffilmark{4}}
\altaffiltext{1}{Faculty of Education, Kagawa University, Saiwai-Cho 1-1, Takamatsu, Kagawa 760-8522}
\email{matsu@ed.kagawa-u.ac.jp}
\altaffiltext{2}{Hiroshima Astrophysical Science Center, Hiroshima University, 
1-3-1 Kagamiyama, Higashi-Hiroshima, Hiroshima 739-8526}
\altaffiltext{3}{Koyama Astronomical Observatory, Kyoto Sangyo University, Motoyama, Kamigamo, Kita-ku, Kyoto 603-8555}
\altaffiltext{4}{Astronomical Institute, Graduate School of Science, Tohoku University, Aramaki, Aoba-ku, Sendai 980-8578}


\maketitle

\begin{abstract}
We investigate the efficiency of interstellar polarization $p_\lambda/A_\lambda$, 
where $p_\lambda$ is the fractional linear polarization and 
$A_\lambda$ is extinction,  
in 14 lines of sight as a function of wavelength $\lambda$. 
We have used the data of lines of sight to the Pleiades cluster obtained 
with the low-dispersion spectropolarimeter HBS as well as those in literature.
It is found that the polarization efficiency $p_\lambda/A_\lambda$ is 
proportional to $\exp(-\beta/\lambda)$
in wavelength $\lambda \approx 0.4\sim0.8 \micron$, where $\beta$ is 
a parameter which varies from 0.5 to 1.2 $\micron$.  
We find that $\beta$ is negatively correlated with the dust temperature 
deduced from infrared data by Schlegel et al.,   
suggesting that the polarization efficiency is higher 
in short wavelength for higher temperature. 
According to the alignment theory by radiative torques (RATs), 
if the radiation is stronger, 
RATs will make small grains align better, and the polarization efficiency 
will increase in short wavelength. 
Our finding of the correlation between $\beta$ and the temperature is 
consistent with what is expected with the alignment mechanism by RATs. 
\end{abstract}

\section{Introduction}

Observed linear polarization in the light from distant stars, 
i.e., often called as "interstellar polarization", is 
interpreted as a phenomenon of dichroic extinction, and shows that 
interstellar grains are optically anisotropic and aligned,  
although the mechanism of the alignment has been still on debate 
(\cite{L07}, for a recent review).
The alignment of grains had been explained with the paramagnetic relaxation 
of thermally spinning grains that obtain angular momentum 
by collisions with gas particles (\cite{DG51}, 
hereafter DG).  However, the DG mechanism is not efficient,  
and it cannot explain the interstellar polarization quantitatively. 
For more efficient alignment, 
\citet{Purcell79} assumed a spin-up of grain 
by the ejection of molecular hydrogen from grain surface  
(the "pinwheel mechanism"), 
though there still remain problems in quantitative explanations 
\citep{LD99,RL99}.

\citet{DM76} first pointed out that irregularly 
shaped grains that have "helicity" can spin up by radiative torques 
(hereafter RATs).  
More recently, \citet{DW96} 
showed that RATs are very effective to align grains. 
Since magnetic moments within rotating grains are induced by the Barnett effect, 
the grains precess around the magnetic field, and 
the direction of alignment is usually parallel to 
the interstellar magnetic field (e.g. \cite{DW97}; \cite{LH07}),  
i.e., the same direction as that by the DG mechanism.  
The alignment by RATs can be more efficient 
if grains have superparamagnetic inclusions \citep{LH08} or 
if the pinwheel mechanism is working with RATs \citep{HL09}.


The efficiency of the RATs alignment varies with strength and spectral energy 
distribution of the radiation field, and thus the size of aligned grains 
should vary accordingly (\cite{DW96}; \cite{ChoLaz07}). 
Observationally, the maximum wavelength $\lambda_{\rm max}$ of polarization 
in dark clouds was shown to be correlated with extinction $A_{\rm V}$ 
in the V-band \citep{Whittet01,AP07}. 
\citet{AP10} showed that grain alignment is enhanced by the stellar radiation 
in the vicinity of a young star HD 97300 in the Chamaeleon I cloud. 
For stars in the Taurus dark cloud,  
\citet{Whittet08} showed that 
the polarization efficiency $p_{\rm K}/\tau_{\rm K}$ in the K-band, 
where $\tau_{\rm K}$ is optical depth, decreases smoothly with $A_{\rm V}$ 
beyond the region where ice mantle feature was detected.  
This suggests that the alignment efficiency is not directly 
related to the state of grain surface, as is expected by the "pinwheel" 
alignment. 
Those observations suggest that the RATs alignment works in dark clouds and 
star forming regions. 
However, it is still not clear whether RATs alignment 
works or not in more diffuse clouds. 

The Pleiades cluster      
is associated with the diffuse reflection nebula, 
where grain alignment may be enhanced by strong stellar radiation 
if the alignment by RATs works.  
This motivated us to observe polarization in the lines of sight to stars 
in the Pleiades cluster with the low-dispersion spectropolarimeter HBS 
\citep{K99}. 
In this Letter, using our polarimetric data and 
those available from literature, 
we investigate correlations between polarization quantities 
and dust temperature, 
because such correlations may be expected from the RATs alignment theory.


%

\section{Observations and Data Reduction}

We observed 8 stars in the Pleiades cluster with the low-dispersion 
spectropolarimeter HBS \citep{K99} attached to the 1.88m telescope 
in Okayama Astrophysical Observatory, from October 30 to November 13 in 2008, 
and from January 13 to 19 in 2009. 
HBS has a superachromatic half-wave plate and a quartz Wollaston prism, 
and it can measure linear polarization over a wide range of 
the optical region, from 0.4 to 0.8$\micron$. 
We used a slit of 0.2mm width, yielding a wavelength resolution of 
$\sim0.006\micron$. 
After binning, we have obtained spectropolarimetric data with resolution of 
$0.02\micron$, and also synthesized V-band data. 
A unit of the observing sequence consists of successive integrations 
at $ 0^\circ$, \ensuremath {22\rlap {.}{}^{\mathrm {\circ }}5}, $ 45^\circ$, 
and \ensuremath {67\rlap {.}{}^{\mathrm {\circ }}5} position angles of 
the half-wave plate. 
The exposure time per object was from 1800 to 17600 sec., 
depending on the brightness of object and also on weather conditions.

The instrumental polarization was evaluated with nonpolarized standard stars,   
i.e., HD 432, HD 21447, HD 95418, HD 198149, and HD 210027 in 2008, and 
HD 21447, HD 20630, HD 95418, HD 114710 and HD 142373 in 2009  
(\cite{K99} and references therein). 
The standard deviations of the fractional Stokes parameters 
for those standard stars were 0.02-0.07$\%$ in the synthesized V-band. 
Since these values are larger than those expected from photon noise, 
i.e. 0.01-0.03$\%$ for each measurement, they can be interpreted as 
instrumental stability in polarization measurement. 
The photon noise for program stars was also small, $\sim0.01\%$.  
We thus consider that the accuracy in our polarimetry is mainly limited 
by the instrumental stability, and estimated the error for fractional 
polarization $p_{\rm V}$ to be 0.04$\%$, which is a mean value of 
standard deviations for nonpolarized standard stars.

The position angle $\theta_{\rm V}$ of linear polarization in the V-band 
was calibrated with polarized standard stars, i.e. HD 7927, HD 43384, 
and HD 204827 in both observational runs. 
Since HD 43384 shows small temporal variation in polarization 
(\cite{HSU82}; \cite{MSK98}), we considered it as a secondary 
standard, though no significant deviations were found 
from the tabulated value of $\theta_{\rm V}=170^\circ$ in literature. 
We thus calibrated $\theta_{\rm V}$ with the accuracy of 
$\sim0.5\rlap{.}{}^{\mathrm {\circ }}$

From the observation of nonpolarized standard stars through 
a Glan-Taylor prism, we estimated the instrumental depolarization
to be 0.95-0.99 depending on wavelength,  and used it for calibration. 
The position angle $\theta_{\lambda}$ in wavelength 
$\lambda=0.4\sim0.8\micron$ was calibrated 
with reference to $\theta_{\rm V}$, with the same observation.

\section{Results and Discussion}
\subsection{Fractional Polarization and Position Angle}

\begin{figure}
  \begin{center}
    \FigureFile(80mm,135mm){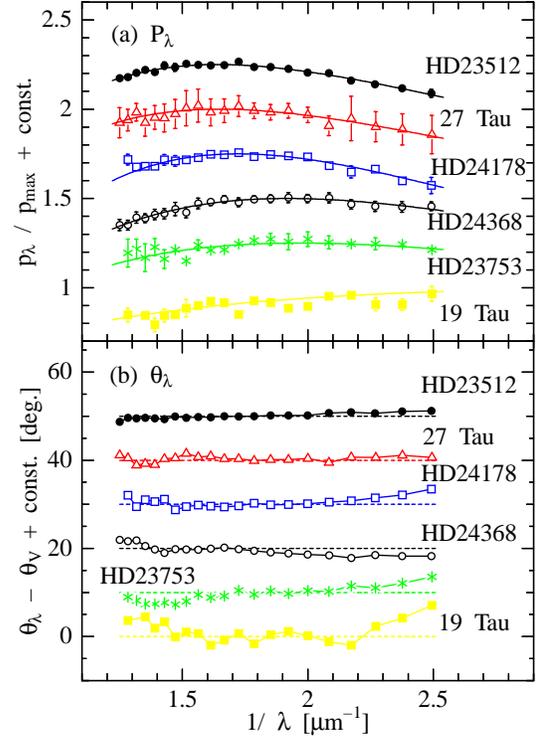}
  \end{center}
  \caption{Wavelength dependence of (a) fractional polarization $p_\lambda$ 
and (b) position angle $\theta_\lambda$ for the Pleiades stars.
Solid lines in (a) show results of fitting by equation (\ref{eq_ser}), and 
dashed lines in (b) the values of $\theta_{\rm V}$. 
Data of each object are moved vertically in steps of 0.25 in (a) and $10^\circ$ in (b).
}\label{fig1}
\end{figure}

The observed fractional polarization $p_\lambda$ and position angle 
$\theta_\lambda$ are shown in Figure \ref{fig1}, 
except for HD 23985 and HD 24118 
which show low polarization $\lesssim0.1\%$. 
We assume an empirical formula for $p_\lambda$ by \citet{Ser75}:
\begin{equation}
p_\lambda = p_{\rm max}\exp(-K\ln^2(\lambda/\lambda_{\rm max})), \label{eq_ser}
\end{equation}
where $p_{\rm max}$ is maximum polarization, 
$K$ is a parameter that determines width of the curve, 
and $\lambda_{\rm max}$ is the wavelength at $p=p_{\rm max}$.
Those derived values of $p_{\rm max}$, $\lambda_{\rm max}$, and $K$ are tabulated 
in Table \ref{tab1}.
Figure \ref{fig1}a shows that $p_\lambda$ is well expressed 
with equation (\ref{eq_ser}). The position angle $\theta_\lambda$ is almost 
constant, though $\theta_\lambda$ of 19 Tau varies with $\lambda$ 
in short wavelength $1/\lambda \gtrsim 2.3\micron^{-1}$ (Figure \ref{fig1}b). 

\begin{table*}
  \caption{Polarization of Stars in the Pleiades Cluster}\label{tab1}
  \begin{center}
    \begin{tabular}{lccrrrcccccccc}
     \hline
Name            & $p_{\rm V}$\footnotemark[$*$] & $p_{\rm max}$\footnotemark[$*$] & $\lambda_{\rm max}$\footnotemark[$\dagger$] & $K$\footnotemark[$\dagger$] &  $\theta_{\rm V}$\footnotemark[$\dagger$] & Method & $R_{\rm V}$ & $A_{\rm V}$ &  $\alpha$ & $\beta$     & $T_{\rm dust}$  \\
\ (Sp.Type)     & [\%]                          & [\%]                            & $[\micron]$                                 &                             &   [deg]                                   &        &             & [mag]       &  [\%/mag] & $[\micron]$ & [K]             \\
      \hline
19 Tau           & 0.26 & 0.28 & 0.36     & 0.26     & 140.2    & 1 & 2.74$\pm.57$  & 0.13 & 2.1$\pm$0.7 &  0.72$\pm$.01 & 20.1  \\
\ \ \ \ \ (B5IV) & -    & -    & $\pm.11$ & $\pm.18$ & $\pm$4.4 & 2 & 3.20$\pm.17$  & 0.13 & 2.1$\pm$0.7 &  0.64$\pm$.01 & -     \\
27 Tau           & 0.34 & 0.35 & 0.61     & 0.91     & 112.6    & 2 & 3.20$\pm.17$  & 0.09\footnotemark[$\ddagger$] & 4.1$\pm$0.8 &  0.94$\pm$.08 & 19.7  \\
\ \ \ \ \ (B8III) & -   & -    & $\pm.03$ & $\pm.45$ & $\pm$3.4 & - & -             & -    & -           &  -            & -     \\
HD23512          & 2.34 & 2.38 & 0.61     & 1.06     & 27.2     & 1 & 3.48$\pm.11$  & 1.15 & 2.09$\pm$.07 & 0.82$\pm$.02 & 20.1  \\
\ \ \ \ \ (A0V)  & -    & -    & $\pm.00$ & $\pm.07$ & $\pm$0.5 & 2 & 3.20$\pm.17$  & 1.15 & 2.12$\pm$.07 & 0.94$\pm$.06 & -     \\
HD23753          & 0.27 & 0.27 & 0.51     & 0.57     & 104.8    & 1 & 3.17$\pm.76$  & 0.09 & 3.1$\pm$1.6  & 0.65$\pm$.14 & 19.3  \\
\ \ \ \ \ (B8V)  & -    & -    & $\pm.01$ & $\pm.16$ & $\pm$4.2 & 2 & 3.20$\pm.17$  & 0.09 & 3.1$\pm$1.6  & 0.67$\pm$.04 & -     \\
HD24178          & 0.50 & 0.50 & 0.58     & 1.38     & 128.2    & 1 & 2.57$\pm.17$  & 0.39 & 1.33$\pm$.13 & 1.00$\pm$.05 & 17.5  \\
\ \ \ \ \ (A0)   & -    & -    & $\pm.00$ & $\pm.13$ & $\pm$2.3 & 2 & 3.20$\pm.17$  & 0.39 & 1.33$\pm$.13 & 0.88$\pm$.06 & -     \\
HD24368          & 0.61 & 0.61 & 0.52     & 0.92     & 95.1     & 1 & 5.16$\pm1.43$ & 0.28 & 2.22$\pm$.31 & 0.45${<-.03 \atop +.15}$ & 17.5  \\
\ \ \ \ \ (A2V)  & -    & -    & $\pm.01$ & $\pm.15$ & $\pm$1.9 & 2 & 3.20$\pm.17$  & 0.28 & 2.28$\pm$.32 & 0.74$\pm$.05 & -     \\
      \hline
   \multicolumn{12}{@{}l@{}}{\hbox to 0pt{\parbox{180mm}{\footnotesize
      \footnotemark[$*$] Errors of $p_{\rm V}$ and $p_{\rm max}$ are estimated to be 0.04\%. See text for details.
      \par\noindent
      \footnotemark[$\dagger$] Errors of $\lambda_{\rm max}$, $K$, and $\theta_{\rm V}$ are written in the 2nd line for each object.
      \par\noindent
      \footnotemark[$\ddagger$] Deduced from the color excess by \citet{Crawford76}, and its error is 0.02 mag. 
                               Other values of $A_{\rm V}$ are derived from 2MASS data, having errors of 0.04 mag. 
      \par\noindent
    }\hss}}

    \end{tabular}
  \end{center}
\end{table*}

Since the scattered light is often strongly polarized, 
it may affect the polarimetric results of nebulous objects, 
if it is not subtracted properly,   
and/or if the objects are surrounded by optically thick cloud, 
e.g. young stellar objects such as R Mon \citep{MSK99}. 
However, the brightness of nebulosity around the stars 
in the Pleiades is not intensive compared with the stellar light, 
typically $\sim 20$ mag/arcsec$^2$ in the B or V-band. 
We thus expect that the effect of the nebulosity is subtracted 
with the sky background in the reduction process. 
In addition, the spectral types of the stars are normal. 
Thus the observed polarization is expected to be mainly 
foreground interstellar in origin.
Nevertheless, 19 Tau may have a component of non-interstellar origin, 
because the position angle is variable:  
$\theta_{\rm V}$ was $114\degree$ in \citet{Mark77} and 
in \citet{Breger86},
but it is $140\pm4\degree$ in our observation (Table \ref{tab1}). 
We thus exclude 19 Tau in the following discussion.

\subsection{Polarization Efficiency}

The $A_{\rm V}$-dependence of $\lambda_{\rm max}$ in dark clouds 
suggests that the alignment of grains is induced by RATs (see Section 1).                        
However, \citet{AP07} noticed that 
the extrapolated value of $\lambda_{\rm max}$ at $A_{\rm V}=0$ 
in each cloud is correlated with the mean value of 
the ratio of total to selective extinction  
$R_{\rm V}$, where $R_{\rm V} \equiv A_{\rm V}/E_{\rm B-V}$  
and $E_{\rm B-V}$ is the color excess for $B-V$.  
Since $R_{\rm V}$ characterizes the extinction,  
this correlation means that $\lambda_{\rm max}$ depends not only 
on the alignment,  
but also on the size of total, i.e. aligned and nonaligned, grains. 
We thus use another quantity less affected by the variation of grain size. 

Compared with $\lambda_{\rm max}$,  
the polarization efficiency $p_{\lambda}/A_{\lambda}$ should be less 
dependent on the variation of grain size, 
because such variation will be canceled in $p_{\lambda}/A_{\lambda}$. 
We thus explore the observed properties of $p_{\lambda}/A_{\lambda}$, 
expecting to obtain information on alignment. 
It should be noted, however, that \citet{Vosh08} and \citet{Das10} 
showed that $p_{\lambda}/A_{\lambda}$ depends on grain size, shape, material, and other parameters. 
It would be possible to examine the properties of $p_{\lambda}/A_{\lambda}$ 
in more detail with using light scattering calculations 
(e.g. Matsumura \& Seki 1991, 1996; \cite{MB09}), 
but it is beyond the scope of this Letter.

To evaluate extinction $A_\lambda$,  we have used two methods:

{\it Method 1}: On the assumption that the $\lambda$-dependence of $A_\lambda$ 
is determined by $R_{\rm V}$ and scaled by $A_{\rm V}$ \citep{CCM89}, 
we evaluate $A_\lambda$ with interpolating the data of $A_\lambda/E_{\rm B-V}$ 
for $R_{\rm V}=2.1\sim5.5$ tabulated in \citet{F04}.  
The values of $A_{\rm V}$ and $R_{\rm V}$ are calculated 
by the formulae $A_{\rm V}=1.1E_{\rm V-K}$ and 
$R_{\rm V}=1.1E_{\rm V-K}/E_{\rm B-V}$ \citep{WvB80}, respectively, 
where $E_{\rm V-K}$ is the color excess for $V-K$. 
We use $B$ and $V$ magnitudes in the Simbad database, 
while for the $K$ band, we transform $K_{\rm S}$ magnitude 
in the Two Micron All Sky Survey (2MASS) into the $K$ magnitude 
in the system of \citet{Koo83} with a formula by \citet{Carpenter01}.  
We refer to \citet{Fitz70} and \citet{Koo83} for 
the intrinsic colors of $B-V$ and $V-K$, respectively.

Since the errors of $R_{\rm V}$ derived with Method 1 are large 
for some stars (Table \ref{tab1}), 
we also use another method as below (Method 2).
The error of $K_{\rm S}$ for 27 Tau was particularly large, 
$\sim0.3$ mag., we could not obtain reliable results, and 
excluded 27 Tau in the following discussion. 

{\it Method 2}:  We assume that the extinction properties are not variable 
within the Pleiades cluster, and apply the extinction curve for HD 23512 
to other stars, scaling it by the value of $A_{\rm V}$ 
deduced with Method 1. 
The extinction curve for HD 23152 is reduced by \citet{FM07}, 
and most reliable among the Pleiades stars. 


We explore not only the Pleiades stars, but also the stars 
for which polarization and extinction data are available from literature.
We have used the polarimetric data of various stars by Weitenbeck (1999, 2004).  
The data of HD 29647 \citep{Whittet01} and HD 38087 \citep{Ser75} are 
used in addition to those from \citet{Weiten99}.
Extinction data for those stars are cited from \citet{FM07}. 
Also used are the data for high latitude clouds MBM 30 and MBM 20 (LDN 1642) 
by \citet{SM96}. 

\begin{figure}
  \begin{center}
    \FigureFile(80mm,135mm){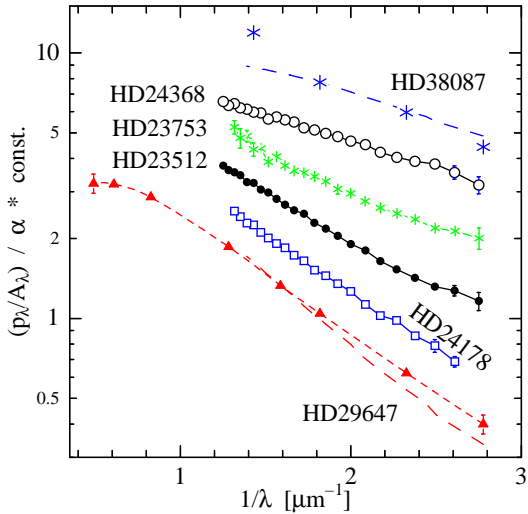}
  \end{center}
  \caption{
Wavelength dependence of polarization efficiency $p_\lambda/A_\lambda$ 
deduced with Method 1 (See text). 
Asterisks show the results of HD 38087 \citep{Ser75} and 
filled triangles HD 29647 \citep{Whittet01},   
while other symbols are for the Pleiades stars as in Figure \ref{fig1}.
Long dashed lines are the results by \citet{Weiten99}. 
The data are normalized by $\alpha$ (see equation (\ref{eq_fit})), 
and moved vertically in steps of factor 1.5. 
  }\label{fig2}
\end{figure}

Figure \ref{fig2} shows the $\lambda$-dependence of $p_\lambda/A_\lambda$,
which is derived with Method 1.   
The values of log($p_\lambda/A_\lambda$) decrease linearly 
with inverse wavelength $1/\lambda$, 
though slight deviations from the linear relation are found. 
We thus  make linear fitting in $\lambda=0.4\sim0.8\micron$ with the equation: 
\begin{equation}
\ln(p_\lambda/A_\lambda) = \ln\alpha - \beta(1/\lambda - 1/0.55\micron),  \label{eq_fit}
\end{equation}
where $\lambda$ is in $\micron$, and 
$\alpha$ and $\beta$ are parameters and tabulated in Tables \ref{tab1} 
and \ref{tab2}. 

\begin{table}
  \caption{Polarization Properties deduced from Literature.}\label{tab2}
  \begin{center}
    \begin{tabular}{lccccc}
     \hline
HD or SAO  & $\alpha$  &  $\beta$    & $T_{\rm dust}$ & Ref.\footnotemark[$*$] \\
(Sp.Type)  & [\%/mag]  & $[\micron]$ & [K]            &       \\
     \hline
 14889 (K0)\footnotemark[$\dagger$] &  2.12$\pm$.21 &  0.46$\pm$.05 & 18.0 & (1) \\
 29647 (B8III) &  0.65$\pm$.02 &  1.06$\pm$.05 & 15.0 & (5) \\
 ---           &  0.62$\pm$.03 &  1.25$\pm$.08 & ---  & (3) \\
 30675 (B3V)   &  2.70$\pm$.13 &  0.83$\pm$.07 & 16.0 & (5) \\
 37367 (B2IV-V) & 0.74$\pm$.04 &  0.76$\pm$.04 & 15.8 & (4) \\
 38087 (B5V)   &  1.47$\pm$.07 &  0.40$\pm$.06 & 21.0 & (3) \\
 ---           &  1.51$\pm$.07 &  0.51$\pm$.03 & ---  & (2) \\
192001 (O9.5IV) & 0.84$\pm$.04 &  0.92$\pm$.05 & 19.8 & (4) \\
193322 (O9V)   &  1.57$\pm$.10 &  0.91$\pm$.08 & 20.4 & (4) \\
210121 (B3V)   &  1.76$\pm$.16 &  0.79$\pm$.11 & 17.4 & (4) \\
216532 (O8.5V) &  0.87$\pm$.02 &  0.74$\pm$.04 & 18.9 & (4) \\
S149760 (K5)\footnotemark[$\dagger$] &  2.06$\pm$.21 &  1.03$\pm$.13 & 16.4 & (1) \\
      \hline
   \multicolumn{5}{@{}l@{}}{\hbox to 0pt{\parbox{85mm}{\footnotesize
      \footnotemark[$*$] References:
	(1): \citet{SM96}, (2): \citet{Ser75}, (3): \citet{Weiten99}, 
	(4): \citet{Weiten04}, (5): \citet{Whittet01}.
        The data in the R-band of HD 38087 in \citet{Ser75} is not used for fitting.
      \par\noindent
      \footnotemark[$\dagger$] Luminosity class is assumed as V, and the values of 
          ($R_{\rm V}$, $A_{\rm V}$) are estimated to be  
          (3.08$\pm$.20, 1.17$\pm$.11) for HD 14889,  and 
          (2.74$\pm$.54, 1.15$\pm$.12) for SAO 149760. 
    }\hss}}
    \end{tabular}
  \end{center}
\end{table}


\subsection{Polarization Efficiency and Dust Temperature}

To discuss the correlations between the polarization properties and 
dust temperature,  
we use the temperature $T_{\rm dust}$ by \citet{SFD98}. 
They deduced $T_{\rm dust}$ from COBE and IRAS data, 
on the assumption of $\lambda^{-2}$ emissivity of grains in the infrared. 
Their data is homogeneous all over the sky, 
and thus suitable for our study that contains not only the lines of sight 
to Pleiades stars, but also those to other objects. 

\begin{figure}
  \begin{center}
    \FigureFile(80mm,110mm){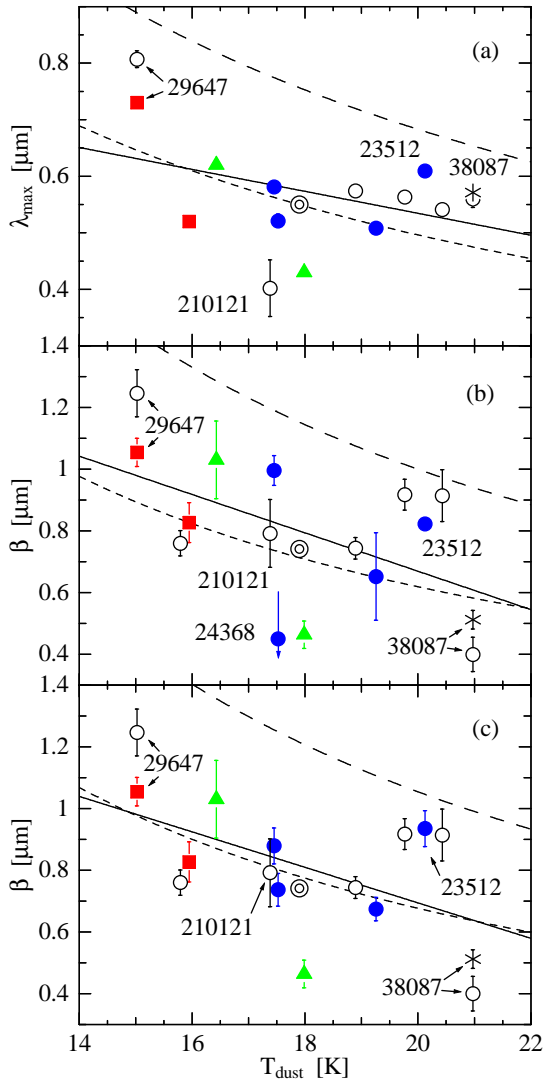}
  \end{center}
  \caption{(a) Correlation between $T_{\rm dust}$ and $\lambda_{\rm max}$. 
Filled circles are the Pleiades stars, open circles the results by 
Weitenbeck (1999, 2004),  filled squares those by \citet{Whittet01}, 
and filled triangles those by \citet{SM96}.
Double circle is the average of interstellar medium. 
Result of linear fitting for 14 stars is drawn as solid line. 
Short dashed line presents a model prediction for $a'=a_{\rm lower}$,
while long dashed line $a'=2a_{\rm lower}$.
See text for details. 
(b) Same as (a) but for $T_{\rm dust}$ and $\beta$ with Method 1. 
(c) Same as (b) but with Method 2.
  }\label{fig3}
\end{figure}

Between $\lambda_{\rm max}$ and $T_{\rm dust}$, 
we find a weak correlation in Figure \ref{fig3}a,  
and the correlation coefficient $r$ is $-0.30$. 
The correlations between $\beta$ and $T_{\rm dust}$ are much better, 
i.e., $r=-0.54$ (Method 1, Figure \ref{fig3}b) and 
$r=-0.57$ (Method 2, Figure \ref{fig3}c).
It is remarkable that the relative positions of HD 210121 and HD 38087 
are different between in Figure \ref{fig3}a and in Figure \ref{fig3}bc. 
This is caused by the different values of $R_{\rm V}$, i.e., 
$R_{\rm V}$ of HD 210121 is smaller ($=2.0$, \cite{FM07}), 
and that of HD 38087 is larger ($=5.8$, \cite{FM07}) 
than other objects ($R_{\rm V}\sim3$).  

We finally discuss the above mentioned correlations  
on the basis of the RATs alignment theory.  
Using their equation (5) in \citet{ChoLaz07} and equation (67) in \citet{DW96}, 
with typical values for physical quantities in interstellar space 
tabulated in Table 2 of \citet{LH07}, 
we can express the smallest size $a_{\rm lower}$ of aligned grains as 
\begin{equation}
 (a_{\rm lower}/1\micron) = 0.089 \times (T_{\rm dust}/18{\rm K})^{-2}, 
                                                             \label{eq_ChoLaz} 
\end{equation}
where we assume that the efficiency $Q_\Gamma$ for RATs is 0.1, 
and that the emissivity of grains is  $\propto \lambda^{-2}$. 

On the size distribution of aligned grains, 
\citet{Mathis86} showed that the observed polarization $p_\lambda$ can be 
reproduced if the fraction $(1-\exp(-(a/a')^3))$ of grains with radius $a$ 
are aligned, where $a'$ is a parameter for typical size of smallest 
aligned grains. \citet{Mathis86} then obtained the equation: 
\begin{equation} 
(a'/1\micron) = 0.327 \times (\lambda_{\rm max}/1\micron)^{2.17}. 
                                                           \label{wmax_a'} 
\end{equation} 
If we assume $a'=a_{\rm lower}$ (or $a'=2a_{\rm lower}$), 
we can relate $\lambda_{\rm max}$ and $T_{\rm dust}$
with equations (\ref{eq_ChoLaz}) and (\ref{wmax_a'}), 
and draw the short dashed (or long dashed) line in Figure \ref{fig3}a.

For the relation between $\beta$ and $\lambda_{\rm max}$, we obtain 
\begin{equation} 
 (\beta/1\micron)=1.79 \times (\lambda_{\rm max}/1\micron)^{1.39},   
\end{equation} 
with using equation (\ref{eq_ser}) and extinction curve for $R_{\rm V}=3.1$ 
\citep{F04}. 
We then write $\beta$ as a function of $T_{\rm dust}$, 
and draw short and long dashed lines, 
for $a'=a_{\rm lower}$ and $a'=2a_{\rm lower}$, respectively,  
in Figures \ref{fig3}b and \ref{fig3}c. 

Those lines in Figure \ref{fig3} seem to follow the observations well, i.e., 
the correlation between $\lambda_{\rm max}$ 
and $T_{\rm dust}$, and that between $\beta$ and $T_{\rm dust}$ 
can be explained by the RATs alignment theory. 
Since those data contain regions of various temperature, 
i.e., the Taurus dark cloud, reflection nebulae, etc.,  
our results suggest that the alignment by RATs is ubiquitous 
in the interstellar space. 

\bigskip
This work was supported by the Thesis Supporting Program 
at Okayama Astrophysical Observatory of NAOJ, NINS (No.08A-S01, 
P.I. was Youko Kameura), 
and by the Kagawa University Specially Promoted Research Fund (FY2008). 
We are grateful to the staff members at Okayama
Astrophysical Observatory for their support during the observations. 
This work has made use of the SIMBAD database, 
operated at CDS, Strasbourg, France.


\begin{thebibliography}{}

\bibitem[Andersson
\& Potter(2007)]{AP07} Andersson, B.-G., \& Potter, S.~B.\ 2007, \apj, 665, 369
\bibitem[Andersson
\& Potter(2010)]{AP10} Andersson, B.-G., \& Potter, S.~B.\ 2010, \apj, 720, 1045
\bibitem[Breger(1986)]{Breger86} Breger, M.\ 1986, \apj, 309, 311 
\bibitem[Cardelli et al.(1989)]{CCM89} Cardelli, J.~A., 
Clayton, G.~C., \& Mathis, J.~S.\ 1989, \apj, 345, 245 
\bibitem[Carpenter(2001)]{Carpenter01} Carpenter, J.~M.\ 2001, \aj, 
121, 2851 
\bibitem[Cho 
\& Lazarian(2007)]{ChoLaz07} Cho, J., \& Lazarian, A.\ 2007, \apj, 669, 1085
\bibitem[Crawford 
\& Perry(1976)]{Crawford76} Crawford, D.~L., \& Perry, C.~L.\ 1976, \aj, 81, 419 
\bibitem[Das et al.(2010)]{Das10} Das, H.~K., Voshchinnikov, 
 N.~V., \& Il'in, V.~B.\ 2010, \mnras, 404, 265 
\bibitem[Davis
\& Greenstein(1951)]{DG51} Davis, L., Jr., \& Greenstein, J.~L.\ 1951, \apj, 114,
206
\bibitem[Dolginov
\& Mitrofanov(1976)]{DM76} Dolginov, A.~Z., \& Mitrofanov, I.~G.\ 1976, \apss, 43, 291
\bibitem[Draine
\& Weingartner(1996)]{DW96} Draine, B.~T., \& Weingartner, J.~C.\ 1996, \apj, 470, 551
\bibitem[Draine
\& Weingartner(1997)]{DW97} Draine, B.~T., \& Weingartner, J.~C.\ 1997, \apj, 480, 633
\bibitem[Fitzgerald(1970)]{Fitz70} Fitzgerald, M.~P.\ 1970, \aap, 4, 234 
\bibitem[Fitzpatrick(2004)]{F04} Fitzpatrick, E.~L.\ 2004,
Astrophysics of Dust, 309, 33
\bibitem[Fitzpatrick
\& Massa(2007)]{FM07} Fitzpatrick, E.~L., \& Massa, D.\ 2007, \apj, 663, 320
\bibitem[Hoang
\& Lazarian(2009)]{HL09} Hoang, T., \& Lazarian, A.\ 2009, \apj, 695, 1457
\bibitem[Hsu 
\& Breger(1982)]{HSU82} Hsu, J.-C., \& Breger, M.\ 1982, \apj, 262, 732 
\bibitem[Kawabata et al.(1999)]{K99} Kawabata, K.~S., et
al.\ 1999, \pasp, 111, 898
\bibitem[Koornneef(1983)]
{Koo83} Koornneef, J.\ 1983, \aap, 128, 84 
\bibitem[Lazarian(2007)]{L07} Lazarian, A.\ 2007, \jqsrt,
106, 225
\bibitem[Lazarian
\& Draine(1999)]{LD99} Lazarian, A., \& Draine, B.~T.\ 1999, \apjl, 516, L37
\bibitem[Lazarian 
\& Hoang(2007)]{LH07} Lazarian, A., \& Hoang, T.\ 2007, \mnras, 378, 910 
\bibitem[Lazarian
\& Hoang(2008)]{LH08} Lazarian, A., \& Hoang, T.\ 2008, \apjl, 676, L25
\bibitem[Markkanen(1977)]{Mark77} Markkanen, T.\ 1977, \aap, 56, 83 
\bibitem[Mathis(1986)]{Mathis86} Mathis, J.~S.\ 1986, \apj, 308, 281
\bibitem[Matsumura 
\& Bastien(2009)]{MB09} Matsumura, M., \& Bastien, P.\ 2009, \apj, 697, 807 
\bibitem[Matsumura 
\& Seki(1991)]{MS91} Matsumura, M., \& Seki, M.\ 1991, \apss, 176, 283 
\bibitem[Matsumura 
\& Seki(1996)]{MS96} Matsumura, M., \& Seki, M.\ 1996, \apj, 456, 557 
\bibitem[Matsumura et al.(1998)]{MSK98} Matsumura, M., Seki, 
M., \& Kawabata, K.~S.\ 1998, Frontiers Science Series No.~23: 
Pulsating Stars: Recent Developments in Theory and Observation, 107 
\bibitem[Matsumura et al.(1999)]{MSK99} Matsumura, M., Seki,
M., \& Kawabata, K.~S.\ 1999, \aj, 117, 429
\bibitem[Purcell(1979)]{Purcell79} Purcell, E.~M.\ 1979, \apj,
231, 404
\bibitem[Roberge \& Lazarian(1999)]{RL99} Roberge, W.~G., 
\& Lazarian, A.\ 1999, \mnras, 305, 615 
\bibitem[Schlegel et al.(1998)]{SFD98} Schlegel, D.~J.,
Finkbeiner, D.~P., \& Davis, M.\ 1998, \apj, 500, 525
\bibitem[Seki
\& Matsumura(1996)]{SM96} Seki, M., \& Matsumura, M.\ 1996, Polarimetry of the Interstellar Medium, 97, 168
\bibitem[Serkowski et al.(1975)]{Ser75} Serkowski, K.,
Mathewson, D.~S., \& Ford, V.~L.\ 1975, \apj, 196, 261
\bibitem[Voshchinnikov 
  \& Das(2008)]{Vosh08} Voshchinnikov, N.~V., \& Das, H.~K.\ 2008, \jqsrt, 109, 1527 
\bibitem[Weitenbeck(1999)]{Weiten99} Weitenbeck, A.~J.\ 1999,
Acta Astronomica, 49, 59
\bibitem[Weitenbeck(2004)]{Weiten04} Weitenbeck, A.~J.\ 2004,
Acta Astronomica, 54, 87
\bibitem[Whittet et al.(2001)]{Whittet01} Whittet, D.~C.~B.,
Gerakines, P.~A., Hough, J.~H., \& Shenoy, S.~S.\ 2001, \apj, 547, 872
\bibitem[Whittet et al.(2008)]{Whittet08} Whittet, D.~C.~B., 
Hough, J.~H., Lazarian, A., \& Hoang, T.\ 2008, \apj, 674, 304 
\bibitem[Whittet 
\& van Breda(1980)]{WvB80} Whittet, D.~C.~B., \& van Breda, I.~G.\ 1980, \mnras, 192, 467 




\end{thebibliography}
\end{document}